\documentclass[aip,apl,superscriptaddress,graphicx,amsmath,amssymb,reprint,showkeys]{revtex4-1}

\usepackage[breaklinks=true,colorlinks=true,linkcolor=blue,urlcolor=blue,citecolor=blue]{hyperref}
\usepackage{graphicx}
\usepackage{dcolumn}
\usepackage{bm}

\usepackage[utf8]{inputenc}
\usepackage[T1]{fontenc}
\usepackage{mathptmx}
\usepackage{etoolbox}

\usepackage{epstopdf}
\usepackage{siunitx}
\usepackage{upgreek}
\usepackage{csquotes}
\usepackage[version=4]{mhchem}

\makeatletter
\def\@email#1#2{%
	\endgroup
	\patchcmd{\titleblock@produce}
	{\frontmatter@RRAPformat}
	{\frontmatter@RRAPformat{\produce@RRAP{*#1\href{mailto:#2}{#2}}}\frontmatter@RRAPformat}
	{}{}
}%
\newcommand*{\rom}[1]{\expandafter\@slowromancap\romannumeral #1@}
\makeatother

\makeatletter
\let\origsection\section
\renewcommand\section{\@ifstar{\starsection}{\nostarsection}}
\newcommand\starsection[1]
{\sectionprelude\origsection*{#1}\sectionpostlude}
\newcommand\sectionprelude{%
	\vspace{0em}}
\newcommand\sectionpostlude{%
	\vspace{-1em}}
\makeatother

\newcommand{\uproman}[1]{\uppercase\expandafter{\romannumeral#1}}

\begin{document}

\title{Nonlinear erasing of propagating spin-wave pulses in thin-film Ga:YIG} 

\author{D. Breitbach}
\email{dbreitba@rptu.de}
\altaffiliation{D. Breitbach and M. Bechberger contributed equally to this work.}
\author{M. Bechberger}%
\altaffiliation{D. Breitbach and M. Bechberger contributed equally to this work.}
\author{B. Heinz}
\author{A. Hamadeh}
\author{J. Maskill}
\affiliation{Fachbereich Physik and Landesforschungszentrum OPTIMAS, \\Rheinland-Pfälzische Technische Universit{\"a}t Kaiserslautern-Landau, \mbox{D-67663 Kaiserslautern, Germany}}
\author{K. O. Levchenko}
\affiliation{\mbox{Faculty of Physics, University of Vienna, Boltzmanngasse 5, A-1090 Wien, Austria}}
\author{\mbox{B. Lägel}}
\affiliation{Fachbereich Physik and Landesforschungszentrum OPTIMAS, \\Rheinland-Pfälzische Technische Universit{\"a}t Kaiserslautern-Landau, \mbox{D-67663 Kaiserslautern, Germany}}

\author{C. Dubs}
\affiliation{\mbox{INNOVENT e.V. Technologieentwicklung, D-07745 Jena, Germany}}

\author{\mbox{Q. Wang}}
\affiliation{\mbox{School of Physics, Huazhong University of Science and Technology, 430074 Wuhan, China}}

\author{\mbox{R. Verba}}
\affiliation{\mbox{Institute of Magnetism, Kyiv 03142, Ukraine}}

\author{\mbox{P. Pirro}}
\affiliation{Fachbereich Physik and Landesforschungszentrum OPTIMAS, \\Rheinland-Pfälzische Technische Universit{\"a}t Kaiserslautern-Landau, \mbox{D-67663 Kaiserslautern, Germany}}

\date{\today}

\begin{abstract} 
Nonlinear phenomena are key for magnon-based information processing, but the nonlinear interaction between two spin-wave signals requires their spatio-temporal overlap which can be challenging for directional processing devices. Our study focuses on a gallium-substituted yttrium iron garnet film, which exhibits an exchange-dominated dispersion relation and thus provides a particularly broad range of group velocities compared to pure YIG. Using time- and space-resolved Brillouin light scattering spectroscopy, we demonstrate the excitation of time-separated spin-wave pulses at different frequencies from the same source, where the delayed pulse catches up with the previously excited pulse and outruns it due to its higher group velocity. By varying the excitation power of the faster pulse, the outcome can be finely tuned from a linear superposition to a nonlinear interaction of both pulses, resulting in a full attenuation of the slower pulse. Therefore, our findings demonstrate the all-magnonic erasing process of a propagating magnonic signal, which enables the realization of complex temporal logic operations with potential application, e.g., in inhibitory neuromorphic functionalities.
\end{abstract}

\maketitle 

The last decade has seen an increasing interest in alternative computing schemes such as neuromorphic computing\cite{Markovic.2020.NeuromorphicComputing,Torrejon.2017.neuromorphic}, particularly in the context of the growing computational demands and the physical limitations of current semiconductor technologies\cite{Waldrop.2016.Moore}. Contemporary research has focused considerable attention on wave-based logic concepts, which offer additional degrees of freedom, a non-Boolean data carrier, and effects such as interference\cite{Hughes.2019.recurrentneural,Shastri.2021.PhotonicNeuromorphic}. In this context, spin-wave based information processing stands out\cite{Mahmoud.2020.SWComputing, Pirro.2021.Roadmap, Csaba.2017.SWComputing}, not least because spin waves exhibit efficient intrinsic nonlinear effects, fundamental to any logic operation. This feature in conjunction with their low power footprint makes spin waves an efficient data carrier for future information processing schemes such as neuromorphic computing\cite{Pirro.2022.Roadmap,Mahmoud.2021.SWLogic}.

To take advantage of nonlinear effects in the spin-wave domain requires the spatio-temporal overlap, i.e., the superposition, of spin-wave signals, which can be challenging for directional processing with propagating spin-waves. Existing spin-wave logic typically makes use of structural design to achieve superposition, e.g., by using waveguides\cite{Pirro.2021.coherentSW,Fischer.2017.Majoritygate, Khitun.2010, Wang.2018.directionalcoupler} or delay structures such as ring resonators\cite{Wang.2020.Ringresonator} to recouple previous signals. However, spin waves typically show a pronounced dispersion and therefore offer intrinsic means to achieve signal superposition in a more elegant way. Here, we study the superposition and interaction of separate spin-wave signals emitted from the same source with a time delay, by exploiting their difference in group velocity, i.e. their dispersive properties. For this purpose, the ultralow spin-wave damping material yttrium iron garnet\cite{Dubs.2020.LPE,Dubs.2017.LPE,Hillebrands.2010.magnonics} (YIG) is well suited since it provides sufficiently long propagation distances as well as functional nonlinear properties. Moreover, it features a positive nonlinear frequency shift when magnetized out-of-plane, a property that facilitates powerful logic functionalities\cite{Wang.2023.OOPYIG}. However, a drawback of this system is the limited range of group velocities in the wavevector regime accessible via standard means of RF excitation. Furthermore, operation in the out-of-plane magnetization configuration poses a challenge for the optical detection of the spin-wave signals.

Here we circumvent these limitations by employing a Gallium-substituted YIG\cite{Boyle.1997.production,Guigay.1985.production, Hansen.1974.production, Goernert.1975.production, Roschmann.1981.production,Boettcher.2022.GaYIG} film instead. This material exhibits a nearly compensated magnetic configuration, resulting in a low saturation magnetization of $M_\text{S} = \SI{20.2}{\milli\tesla}$, which leads to an almost vanishing dipolar contribution to the spin-wave dispersion. More importantly, Ga:YIG shows a large perpendicular magnetic anisotropy (PMA) and an exchange length which is several times larger compared to pure YIG\cite{Boettcher.2022.GaYIG,Klingler.2015.exchangestiffness}. Therefore, the spin-wave dispersion in this material is exchange-dominated even at low wavevectors and of almost isotropic, quadratic shape, which is analogous to the dispersion of a non-relativistic particle in quantum mechanics. Thus, the group velocity has a linear wavevector dependence and covers a broad range with fine tunability compared to pure YIG. This allows for the excitation of spin-wave signals with a particularly broad difference in group velocities. Furthermore, due to the PMA, the material shows a positive nonlinear frequency shift coefficient\cite{Carmiggelt.2021.GaYIG} when magnetized in the in-plane configuration, easily accessible  by optical means such as Brillouin light scattering spectroscopy (BLS). In this work, we utilize these unique features to demonstrate the superposition of propagating spin-wave signals which are excited by the same source but with a time delay. Furthermore, we study their nonlinear interaction with regard to the functionality of spin-wave erasing - a process where one temporal signal annihilates another.

\begin{figure}[t!]
	\includegraphics{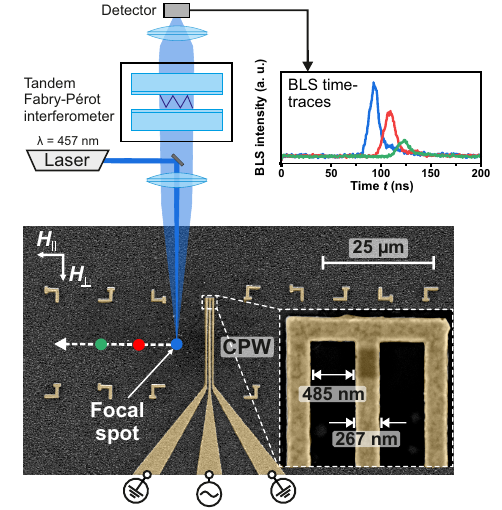}%
	\caption{\label{fig:Setup} Colorized SEM micrograph of the CPW antenna under investigation, placed on top of the Ga:YIG film and a schematic of the applied time-resolved BLS microscope. In the experiment, an in-plane bias magnetic field of $\upmu_{\mathrm{0}}\textit{H}_\text{app} \approx \qty{86}{mT}$ is applied either parallel or perpendicular to the scan direction to magnetize the Ga:YIG in-plane. Timetraces of an exemplary BLS measurement of a propagating spin-wave pulse excited at $f = \SI{1.4}{\giga\hertz}$ are shown to illustrate the measurement process.}
\end{figure}

Figure \ref{fig:Setup} shows the investigated structure and the experimental setup. The magnonic material under study is an \SI{59}{\nano\meter} thin Ga:YIG film, grown by liquid phase epitaxy (LPE)\cite{Dubs.2020.LPE,Dubs.2017.LPE} adjusted for the deposition of sub-100-nm, single crystalline \ce{Y3Fe_{5-x}Ga_{x}O12} films. Its material properties have been thoroughly studied in a previous work\cite{Boettcher.2022.GaYIG}. It is characterized by a strong PMA of $\upmu_{\mathrm{0}}H_\text{u} = \SI{94.1+-0.5}{\milli\tesla}$, an exchange length of $\lambda_\text{ex} = \SI{91.9}{\nano\meter}$, which is substantially larger compared to pure YIG and the low Gilbert-damping parameter of $\alpha = \SI{6.1+-0.6e-4}{}$. Unless stated otherwise, for all presented results an external in-plane magnetic field of $\upmu_{\mathrm{0}}H_\perp = \SI{86\pm2}{\milli\tesla}$ is applied to ensure that the magnetization of the film is aligned in the film plane (see supplementary material, Sec. \rom{1}).\\
On top of the film, a coplanar waveguide (CPW) antenna is structured for spin-wave excitation, composed of Ti/Au (\SI{10}{\nano\meter}/\SI{70}{\nano\meter}). The antenna is fed by a microwave setup consisting of two microwave sources connected to two individual microwave switches which are separately triggered by a pulse generator. The two circuits are then merged by a Wilkinson type power combiner such that the backward isolation is ensured. Using this setup, short microwave pulses of approximately $\tau_\text{rf}\approx \SI{15}{\nano\second}$ are generated to excite a propagating spin-wave packet with a certain carrier frequency. The accessible frequencies are defined by the excitation efficiency of the CPW antenna, which is a function of the wavevector ${k}$ of the excited spin waves (see supplementary material, \mbox{Sec. \rom{2}}).
The experimental investigation of the spin-wave propagation is studied by applying time- and space-resolved BLS\cite{Sebastian.2015.BLS}, focusing a laser beam of $\lambda = \SI{457}{\nano\meter}$ with a laser power of $P = \SI{3.0}{\milli\watt}$ on the sample. After the spin-wave pulse is excited, it propagates away from the CPW antenna. As schematically depicted in \mbox{Fig. \ref{fig:Setup}}, time-resolved BLS measurements are performed at different distances to the antenna, allowing to track the spin-wave pulse in time and space and to extract its group velocity. 

\begin{figure}[t!]
	\includegraphics{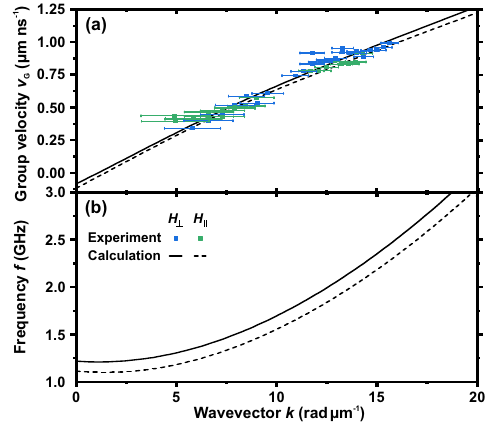}
	\caption{\label{fig:vg} \textbf{(a)} Group velocity as a function of the spin-wave wavevector ${k}$, theoretical curve (black) and experimental data (points). Each datapoint is extracted from a 1D spatial BLS scan, respectively. The pulses were excited using an excitation pulse length of $\tau_{\mathrm{RF}}\approx\SI{10}{\nano\second}$ and varying microwave powers to account for the excitation efficiency of the CPW (see supplementary material, Sec. \rom{3}). The datapoints were converted from excitation frequency to the corresponding wavevector using the dispersion relation shown in panel \textbf{(b)}, calculated according to Ref\cite{Boettcher.2022.GaYIG}. The external field values are $\upmu_{\mathrm{0}}H_\perp = \SI{88.54}{\milli\tesla}$ and $\upmu_{\mathrm{0}}H_\parallel = \SI{85.76}{\milli\tesla}$. }
\end{figure}

The resulting group velocities $v_{\mathrm{g}}$ are shown in \mbox{Fig. \ref{fig:vg}} \textbf{(a)} as a function of the excited spin-wave wavevector $\mathbf{k}$. The data was collected under two field configurations, $\mathbf{k}\perp\mathbf{M}$ and $\mathbf{k}\parallel\mathbf{M}$. In addition, the dispersion relation of both cases is depicted in \mbox{Fig. \ref{fig:vg}} \textbf{(b)}. It can be seen from these two configurations at approximately the same field values ${H}_{\mathrm{app}}$ that the spin-wave properties are highly isotropic, confirming the exchange-dominated character of this system. \mbox{Figure \ref{fig:vg}} also shows the quadratic shape of the dispersion relation $\omega(k)$. This results in the group velocity $v_{\mathrm{g}}=\frac{\partial \omega}{\partial k}$ being almost linearly dependent on the spin-wave wavevector ${k}$, which is in good agreement with the obtained data.
Due to this property, the system exhibits a particularly large range of group velocities in the low-wavevector regime compared to pure YIG, ranging from approximately $v_{\mathrm{g}}= \SI{0}{}$ at $k = \SI{0}{}$ to above $v_{\mathrm{g}}=\SI{1}{\micro\meter\per\nano\second}$ at $k = \SI{15}{\radian\per\micro\meter}$. This property allows for an interesting experiment: spin-wave pulses with strongly different group velocities excited from the same source could still come into superposition even when excited at different times. Due to the high isotropy of the system, the following investigations are carried out in $\mathbf{k}\perp\mathbf{M}$ configuration.\\

\begin{figure}[t!]
	\includegraphics{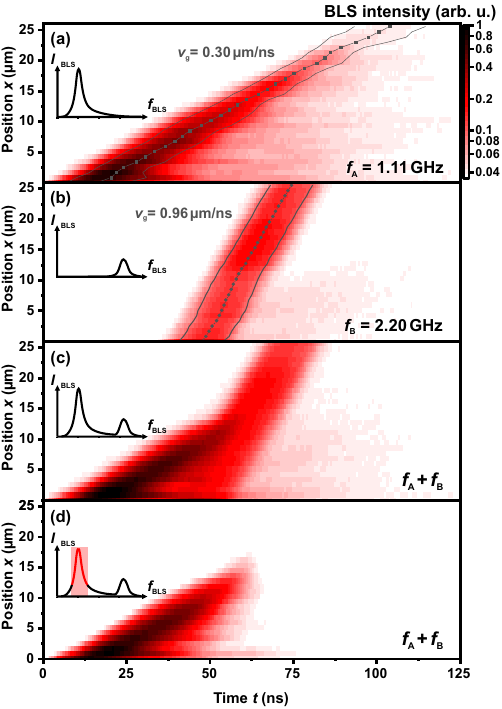}%
	\caption{\label{fig:Superposition_2d} Time-resolved BLS measurements of two short pulses \textbf{A} and \textbf{B} excited at different frequencies $f_{\mathrm{A}} = \SI{1.11}{\giga\hertz}$ and $f_{\mathrm{B}} = \qty{2.2}{\giga\hertz}$ and with a time delay of $\Delta t_\text{0} \approx \qty{32}{ns}$. \textbf{(a)} Only excitation of pulse \textbf{A}. \textbf{(b)} Only excitation of pulse \textbf{B}. \textbf{(c)} Combined excitation of pulses \textbf{A} and \textbf{B} with same parameters as in previous panels. All color plots have the same color scale and normalization. \textbf{(d)} Same as \textbf{(c)} but only BLS frequencies around $f_{\mathrm{A}}$ are shown (marked in red in the inset). Parameters: $P_{\mathrm{A}} = \SI{0}{\decibel m}$ and $P_{\mathrm{B}} = \SI{15}{\decibel m}$.} 
\end{figure}

\mbox{Figure \ref{fig:Superposition_2d} \textbf{(a)}} shows the time- and space-resolved measurement of a spin-wave pulse excited close to the band bottom at $f_{\mathrm{\mathbf{A}}} = \SI{1.11}{\giga\hertz}$, which corresponds to a low wavevector of ${k}_{\mathrm{\mathbf{A}}} \approx \SI{5+-1}{\radian\per\micro\meter}$. On the other hand, \mbox{Fig. \ref{fig:Superposition_2d} \textbf{(b)}} depicts the measurement of a spin-wave pulse that was excited with a frequency of $f_{\mathrm{\mathbf{B}}} = \SI{2.2}{\giga\hertz}$ and a wavevector of ${k}_{\mathrm{\mathbf{B}}}\approx \SI{15+-0.3}{\radian\per\micro\meter}$. It is noted that while the frequency $f_{\mathrm{A}}$ is required to be close to the band bottom, the exact choice of $f_{\mathbf{B}}$ is not crucial to the effect under study, but rather an experimental trade-off between a high group velocity and a feasible excitation efficiency of the CPW antenna (see supplementary material, Sec. \rom{4}). Close to the CPW antenna, the pulses have a duration (FWHM) of $\tau_{\mathbf{A}} = \SI{17}{\nano\second}$ and $\tau_{\mathbf{B}} = \SI{14}{\nano\second}$. The pulse timings and their respective width (FWHM) are marked on the color plot as extracted by fitting the time-dependent data with Gaussian functions. In the following, these two pulses will be referred to as pulse \textbf{A} and \textbf{B}, respectively.\\
It is evident from this measurement that pulse \textbf{A} and \textbf{B} significantly differ in their group velocity. While pulse \textbf{A} with $v_{\mathrm{g}} = \SI{0.30}{\micro\meter\per\nano\second}$ is much slower, pulse \textbf{B} propagates with $v_{\mathrm{g}} = \SI{0.96}{\micro\meter\per\nano\second}$, which is also the main reason for their different decay behavior (see supplementary material, Sec. \rom{5}).

In the following, both pulses are combined in a single experiment while all other parameters are kept constant. The result is shown in \mbox{Fig. \ref{fig:Superposition_2d} \textbf{(c)}} and demonstrates that, even with a time delay of $\Delta t_\text{0} = \SI{32}{\nano\second}$ in their excitation, pulse \textbf{B} catches up with pulse \textbf{A} approximately \SI{14}{\nano\second} after its excitation, leading to a crossing point. However, pulse \textbf{A} vanishes when being passed by the faster pulse \textbf{B}, indicating a direct interaction between both pulses. This becomes evident in \mbox{Fig. \ref{fig:Superposition_2d} \textbf{(d)}}, which shows the combined measurement of both pulses ($f_{\mathrm{A}}+f_{\mathrm{B}}$), but extracted only for BLS frequencies around $f_{\mathrm{A}}$. The fast pulse \textbf{B} provides a moving spatial barrier for spin waves in the low frequency range and erases the slow pulse \textbf{A} including its pulse tail, which would otherwise reach the furthest point of the measurement (see also supplementary material, Sec. \rom{6}). This means that spin-wave signals generated at different times but from the same source can still be brought to superposition and even nonlinearly interact.

\begin{figure}[t!]
	\includegraphics{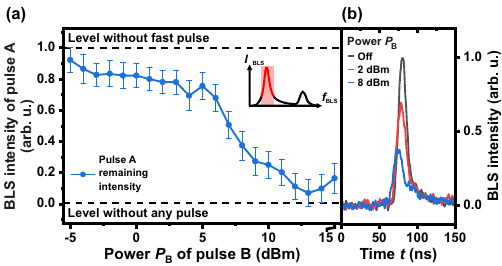}%
	\caption{\label{fig:Erasing} \textbf{(a)} BLS intensity of slow pulse \textbf{A}, extracted for BLS frequencies around $f_{\mathrm{A}}$, as a function of the applied microwave power $P_{\mathrm{B}}$ of fast pulse \textbf{B} (see also supplementary material, Sec. \rom{4}). \textbf{(b)} Timetraces of pulse \textbf{A} extracted for BLS frequencies around $f_{\mathrm{A}}$ for different microwave powers $P_{\mathrm{B}}$. Parameters: $P_{\mathbf{A}} = \SI{0}{\decibel m}$, position: $x = \SI{20.5}{\um}$.}
\end{figure}

\begin{figure*}[t!]
	\includegraphics{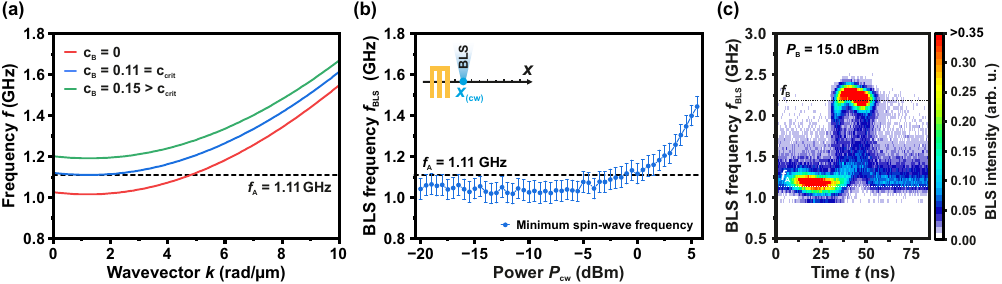}%
	\caption{\label{fig:Nonlinear_Upshift} Effect of the nonlinear frequency cross-shift of the mode $k_\textbf{B} = \SI{15}{\radian\per\micro\meter}$ (not shown) on the linear spin-wave dispersion. \textbf{(a)} Calculated spin-wave dispersion relation for the linear case (red) and cases with increased amplitude $c_\textbf{B}$ at critical (blue) and overcritical level (green). Calculated for $\upmu_{\mathrm{0}}H_{\mathrm{eff}} = \SI{88.5}{mT}$ which includes effects of crystal anisotropy to match the experiment. \textbf{(b)} Measured minimal frequency of the thermal spin-wave signal (BLS) as a function of the applied, continuous microwave power $P_{\mathrm{cw}}$ at $f_{\mathrm{B}} = \SI{2.2}{\giga\hertz}$. Measured at position $x_{\mathrm{(cw)}} = \SI{4.0}{\um}$. \textbf{(c)} Time-resolved BLS measurement of pulses \textbf{A} and \textbf{B} before the pulse crossing at the power levels $P_{\mathrm{A}} = \SI{0}{\decibel m}$ and $P_{\mathrm{B}} = \SI{15}{\decibel m}$, measured at position $x = \SI{4.0}{\um}$.}
\end{figure*}

To study the nonlinear interaction between the pulses, the remaining intensity of pulse \textbf{A} at position $x = \SI{20.5}{\um}$ is considered as a function of the applied microwave power $P_{\mathrm{B}}$ of the fast pulse \textbf{B}, see \mbox{Fig. \ref{fig:Erasing} \textbf{(a)} and \textbf{(b)}}. As $P_{\mathrm{B}}$ increases, the intensity of the slow pulse decreases from being almost unaffected to almost full attenuation. While pulse \textbf{A} retains around 90\% of its original intensity for $P_{\mathrm{B}} = \SI{-5}{\decibel m}$, it is reduced to only 10\% at $P_{\mathrm{B}} = \SI{13}{\decibel m}$. Varying the excitation power of the pulse \textbf{B} therefore enables either an almost linear superposition of both pulses or a nearly complete erasure of the slower pulse \textbf{A}. The results of \mbox{Figs. \ref{fig:Superposition_2d} and \mbox{\ref{fig:Erasing}}} are confirmed by micromagnetic simulations (see supplementary material, \mbox{Sec. \rom{7})}.\\ 

To investigate the physical process behind this mechanism, the effect of the nonlinear frequency shift on the linear dispersion is considered. The four-magnon interaction between two waves with wavevectors $\mathbf{k}$ and $\mathbf{k'}$ results, in particular, in a total shift of the spin-wave frequency of mode $\mathbf{k}$ \cite{Lvov_Book1994},
    \begin{equation}\label{equ:nonlinear_shift}
        \tilde\omega_\mathbf{k} = \omega_\mathbf{k} + T_\mathbf{k} |c_\mathbf{k}|^2 + 2 T_{\mathbf{k}\mathbf{k'}} |c_\mathbf{k'}|^2 \,,
    \end{equation}
where $\omega_\mathbf{k} = 2\pi f_\mathbf{k}$ is the linear spin-wave frequency, $c_\mathbf{k}$ and $c_\mathbf{k'}$ are the spin-wave amplitudes of the two modes, $T_\mathbf{k}$ is the self nonlinear frequency shift coefficient and $T_{\mathbf{k}\mathbf{k'}}$ is the cross nonlinear frequency shift.
Calculations with the Hamiltonian formalism \cite{Patton.2010.Hamiltonian} result in a positive cross-shift of $T_{\mathbf{k}_\mathrm{A}\mathbf{k}_\mathrm{B}} / (2\pi) \approx \SI{3.6}{\giga\hertz}$, as expected for an in-plane magnetized film with high PMA. This contribution is crucial for the comprehension of the experimental results -- pulse \textbf{B} shifts the entire dispersion to higher frequencies, in particular, the bottom of the dispersion, see Fig. \ref{fig:Nonlinear_Upshift} \textbf{(a)}. The dispersion relations shown are calculated according to Eq. (\ref{equ:nonlinear_shift}) using different spin wave amplitudes $c_\textbf{B}$: $f_{\mathbf{k}} = f(k,c_\textbf{k}\rightarrow 0,c_\textbf{B})$. The dispersion in the absence of pulse \textbf{B}, i.e. the linear dispersion, is shown in red. In the presence of pulse \textbf{B}, it is increasingly shifted to higher frequencies for an increasing amplitude $c_\textbf{B}$. When a critical value of $c_\textbf{B}$ = 0.11 is reached, corresponding to about 9 degree magnetization precession angle, the minimal frequency of the dispersion relation is shifted up to $f_\textbf{A} =$ \SI{1.1}{\giga\hertz}. For higher amplitudes $c_\textbf{B}$, this low-frequency mode no longer exists and pulse \textbf{A} becomes nonresonant. Hence, as long as the frequency is conserved, this prevents further propagation of the slow pulse \textbf{A}. This mechanism occurs when the fast pulse passes the slower pulse with a spin-wave amplitude $c_\textbf{B} > c_\text{crit}$, which results in the cut-off of pulse \textbf{A}. Note that according to Eq. (\ref{equ:nonlinear_shift}) this mechanism does not depend on the phase of pulse \textbf{B}, but only on its intensity $|c_\mathbf{B}|^2$.\\
Due to the inaccessibility of absolute spin-wave amplitudes $c_\textbf{k}$ by means of BLS spectroscopy, only qualitative comparisons can be made between the experiment and this calculation. This can be achieved using a continuous microwave excitation at the frequency $f_{\mathrm{B}}$ with increasing excitation amplitude. This allows to investigate the effect of the nonlinear shift on the thermal spin-wave population, i.e. incoherent, thermally excited spin waves that can be detected by BLS at room temperature. Fig. \ref{fig:Nonlinear_Upshift} \textbf{(b)} shows the measured minimal frequency of the thermal spin-wave spectrum as a function of the continuous excitation power. The minimum frequency increases to $f_\textbf{A}$ at a power of $P_\text{cw} \approx\SI{0}{\decibel m}$ and increases even further for higher amplitudes. This experiment demonstrates that using experimentally feasible spin-wave amplitudes, the linear spin-wave frequency can be shifted by several $\SI{100}{\mega\hertz}$.\\ 
Since the previous experiments were done using pulsed RF excitation, the power $P_\textbf{B}$ cannot be directly compared to the continuous power levels in Fig. \ref{fig:Nonlinear_Upshift} \textbf{(b)}. Fig. \ref{fig:Nonlinear_Upshift} \textbf{(c)} shows a frequency- and time-resolved BLS spectrum of the pulses \textbf{A} and \textbf{B} for an excitation power of $P_\textbf{B}$ = \SI{15}{\decibel m} before the pulse crossing. At this power level, pulse \textbf{B} can almost completely erase pulse \textbf{A}, as was shown in Fig. \ref{fig:Erasing}. It is visible that the thermal spin-wave spectrum shifts to frequencies higher than $f_\textbf{A}$ in the presence of pulse \textbf{B}, which prevents further propagation of the slow pulse. This interpretation is in agreement with the power-dependent attenuation in Fig. \ref{fig:Erasing}, as the nonlinear frequency shift is proportional to the spin-wave intensity, which increases with the applied microwave power (see Fig. \ref{fig:Nonlinear_Upshift} \textbf{(b)}). This mechanism occurs when the fast pulse passes the slower one, which results in the cut-off of pulse \textbf{A}. Effectively, the faster pulse is an eraser for a spin-wave signal that was sent earlier.

It remains an open question, however, how the energy of pulse \textbf{A} is dissipated or redistributed during the erasing process. Potential channels for this could include reflection or even scattering perpendicular to the propagation direction, or scattering to the whole spin-wave spectrum as the mode becomes nonresonant. In our current study, no clear signature of such energy transfer could be found. The difficulty to unravel the precise mechanisms lies not only in the two-dimensional extension of the investigated film, but also the wavevector dependent and -limited BLS sensitivity, which can effectively obscure energy scattering to different frequency channels. It is noted that in Fig. \ref{fig:Nonlinear_Upshift} (c), spin-waves at intermediate frequencies occur during the presence of pulse \textbf{B}. However, these are caused by the intense pulse \textbf{B} and are not related to an energy transfer of pulse \textbf{A}, as they also occur in absence of pulse \textbf{A}.
The clarification of this question requires further experiments and extensive simulations as well as studying a confined waveguide.

In conclusion, we demonstrated that temporally separated spin-wave signals excited by the same source can be brought into direct superposition by exploiting a systems intrinsic dispersive properties. We achieved this by employing Ga:YIG, a low-damping material with an exchange-dominated dispersion relation and a high range of excitable group velocities with fine tunability. Utilizing a considerable difference in group velocity, two temporally separated spin-wave pulses can be excited such that the second pulse catches up with the first, bringing them into direct superposition. As a proof-of-principle for the interaction between the pulses, we have demonstrated the tunable nonlinear erasing of the slow spin-wave pulse by the faster pulse. This is mediated by a positive nonlinear frequency shift induced by the fast pulse, which locally shifts the spin-wave dispersion to higher frequencies. This renders the slow pulse nonresonant, hence hindering its further propagation. Our work demonstrates the direct nonlinear interaction of signals from separated temporal inputs which is key to the realization of complex temporal logic operations and is essential for recurrent neuromorphic computing functionalities such as the fading memory. Furthermore, the ability to erase previous information is potentially applicable as an inhibitory component in neuromorphic computing.\newline

See the supplementary material for a additional data on the in-plane magnetization of the Ga:YIG film, on the excitation efficiency of the CPW antenna and for details on the selected power levels for the group velocity measurements. Furthermore, the supplementary material contains details on the choice of excitation frequencies, analysis of the spin-wave decay and additional time-domain data on the erasing process. Lastly, comparative micromagnetic simulations are shown.

\begin{acknowledgements}
This research was funded by the European Research Council within the Starting Grant No. 101042439 "CoSpiN", by the Deutsche Forschungsgemeinschaft (DFG, German Research Foundation) within the Transregional Collaborative Research Center—TRR 173–268565370 “Spin + X” (project B01) and the project 271741898. The authors acknowledge support by the Max Planck Graduate Center with the Johannes Gutenberg-Universität Mainz (MPGC). R.V. acknowledges support by MES of Ukraine (project 0124U000270). K. L. acknowledges the Austrian Science Fund FWF for the support through Grant ESP 526-N "TopMag". Q. W. acknowledges support from the National Key Research and Development Program of China (Grant Nos. 2023YFA1406600) and National Natural Science Foundation of China, the startup grant of Huazhong University of Science and Technology (Grants No. 3034012104). 
\end{acknowledgements}

\section*{Author declarations}
\subsection*{Conflict of Interest}
The authors declare no competing interests.

\section*{Data availability}
The data that support the findings of this study are available from the corresponding author upon reasonable request.

\section*{References}
%

\end{document}